\documentclass[reprint,
showpacs,preprintnumbers,showkeys,
 amsmath,amssymb,
 aps, prc
]{revtex4-1}

\usepackage{graphicx}% Include figure files
\usepackage{float}% zur Verwendung von -H für Bilderplatzierung
\usepackage{dcolumn}% Align table columns on decimal point
\usepackage{bm}% bold math
\begin{document}

\preprint{APS/123-QED}

\title{Measurement of the stellar $^{58}$Ni$(n,\gamma)^{59}$Ni cross section with AMS}% Force line breaks with \\
%\thanks{This work has been funded by The DFG Excellence Cluster Universe and...}%

\author{Peter~Ludwig}
 \email{peter.ludwig@ph.tum.de}
\author{Georg~Rugel} 
 \altaffiliation[Present address: ]{Helmholtz-Zentrum Dresden-Rossendorf, Helmholtz Institute Freiberg for Resource Technology, D-01328 Dresden, Germany}
\author{Iris~Dillmann}
 \email{dillmann@triumf.ca}
 %\altaffiliation[Previous address: ]{Karlsruhe Institute of Technology (KIT), Campus Nord, Karlsruhe, Germany}
 \altaffiliation[Present address: ]{TRIUMF, 4004 Wesbrook Mall, Vancouver BC, and Department of Physics and Astronomy, University of Victoria, Victoria BC, Canada}
\author{Thomas~Faestermann}
\author{Leticia~Fimiani}
\author{Karin~Hain}
 \altaffiliation[Present address: ]{Isotope Research and Nuclear Physics, Faculty of Physics, University of Vienna, Vienna, Austria}
\author{Gunther~Korschinek}
\author{Johannes~Lachner}
 \altaffiliation[Present address: ]{Isotope Research and Nuclear Physics, Faculty of Physics, University of Vienna, Vienna, Austria}
\author{Mikhail~Poutivtsev}

\author{Klaus~Knie}
 \altaffiliation[Present address: ]{FAIR GmbH, D-64291 Darmstadt, Germany}
\affiliation{Physik Department E12 and E15 and Excellence Cluster Universe, Technische Universit\"at M\"unchen, D-85748 Garching, Germany}
  
\author{Michael Heil}
 \altaffiliation[Present address: ]{GSI Helmholtzzentrum f\"ur Schwerionenforschung, D-64291 Darmstadt, Germany}
\author{Franz K\"appeler}
\affiliation{Karlsruhe Institute of Technology (KIT), Campus Nord, Karlsruhe, Germany}

\author{Anton~Wallner}
\affiliation{Isotope Research and Nuclear Physics, Faculty of Physics, University of Vienna, Vienna, Austria\\ and Department of Nuclear Physics, The Australian National University, Canberra, ACT 2601, Australia}

\date{\today}% It is always \today, today,

\begin{abstract}
The $^{58}$Ni$(n,\gamma)^{59}$Ni cross section was measured with a combination of the activation technique and accelerator mass spectrometry (AMS). The neutron activations were performed at the Karlsruhe 3.7~MV Van de Graaff accelerator using the quasi-stellar neutron spectrum at $kT=25$~keV produced by the $^7$Li($p,n$)$^7$Be reaction. The subsequent AMS measurements were carried out at the 14~MV tandem accelerator of the Maier-Leibnitz-Laboratory in Garching using the Gas-filled Analyzing Magnet System (GAMS). Three individual samples were measured, yielding a Maxwellian-averaged cross section at $kT=30$~keV of $\langle\sigma\rangle_{30\text{keV}}$=~30.4 (23)$^{syst}$(9)$^{stat}$~mbarn. This value is slightly lower than two recently published measurements using the time-of-flight (TOF) method, but agrees within the uncertainties. Our new results also resolve the large discrepancy between older TOF measurements and our previous value.

%\begin{description}
%\item[Usage] Secondary publications and information retrieval purposes.
%\item[Structure] You may use the \texttt{description} environment to structure your abstract;
%\end{description}
\end{abstract}

% PACS

\maketitle

%\tableofcontents
%%%%%%%%%%%%%%%%%%%%%%%%%%%%%%%%%%%%%%%%%%%%%%%%%%%%%
\section{Introduction}\label{intro}
%%%%%%%%%%%%%%%%%%%%%%%%%%%%%%%%%%%%%%%%%%%%%%%%%%%%%
$^{58}$Ni is the most abundant stable isotope of nickel (Table~\ref{tab:ni}) and of special interest for astrophysics and reactor technologies. 

Neutron capture on $^{58}$Ni produces the long-lived isotope $^{59}$Ni which has an adopted half-life of $t_{1/2}$= 76(5)~ka \cite{Nish81}. However, recent measurements have shown that this value could be as high as 97(9)~ka \cite{Wall07} .
Since $^{58}$Ni is an important constituent of structure materials in nuclear reactors, neutron activation leads to a potential radiation hazard. In astrophysics it is an important seed nucleus for the "slow neutron capture" ($s$) process, and the half-life of $^{59}$Ni is long enough that under typical conditions no significant branching occurs and the main reaction flow continues via $^{58}$Ni$(n,\gamma)^{59}$Ni$(n,\gamma)^{60}$Ni.

\begin{table}[!b]
\caption{Isotopic \cite{iupac11} and solar abundances (based on meteorites and relative to $A_{\rm{Si}}$= 10$^6$) \cite{AG89} of the stable Ni isotopes. \label{tab:ni}}
\renewcommand{\arraystretch}{1.2} % enlarge line spacing
\begin{ruledtabular}
\begin{tabular}{ccc}
Isotope		& Isotopic abundance [\%]  	& Solar abundance \\ %$A_i$
\hline
$^{58}$Ni	& 68.077 (19)							&  3.36$\times$10$^4$\\
$^{60}$Ni	& 26.223 (15)							&  1.29$\times$10$^4$\\
$^{61}$Ni	& 1.140 (1)								&  5.62$\times$10$^2$\\
$^{62}$Ni	& 3.635 (4)								&  1.79$\times$10$^3$\\
$^{64}$Ni	& 0.926	(2)								&  4.57$\times$10$^2$\\
\hline
$\Sigma_{\rm{Ni}}$ & 100						&  4.93$\times$10$^4$\\
\end{tabular}
\end{ruledtabular}
\end{table}

\subsection{Production of iron-group elements in stars}
Woosley \textit{et al.} \cite{WAC73} demonstrated in 1973 that in massive stars ($M>8~M_\odot$), a superposition of late burning stages of explosive oxygen and silicon burning, provides a good fit to the solar abundances in the mass region $28<A<62$. 
%At the beginning of the Si burning phase the most abundant isotopes are $^{28}$Si and $^{32}$S which are partially photodisintegrated due to the high temperatures to lighter nuclei. The released $\alpha$-particles are then added to heavier nuclei forming the isotopes up to iron-group nuclei in a massive photodisintegration rearrangement. 
The final composition depends on the respective peak temperatures, densities, and the available amount of protons, neutrons, and $\alpha$-particles. 

Explosive silicon burning occurs at $T$$\geq$4~GK and can be subdivided into incomplete burning, complete burning with normal freeze-out, and complete burning with $\alpha$-rich freeze-out. 

Incomplete silicon burning occurs at peak temperatures of $T$= 4$-$5~GK when the temperature is not high enough for nuclear reactions to overcome the bottleneck at the magic shell closure $Z=20$ (Ca). The most abundant burning products are the same as for explosive oxygen burning, but partial leakage can produce iron-group elements. 

Complete Si burning is possible for $T>5$~GK where a full Nuclear Statistical Equilibrium (NSE) is established and iron-group elements like $^{58}$Ni are produced. Complete Si burning with $\alpha$-rich freeze-out occurs at lower densities when the triple $\alpha$-reaction is not fast enough to keep the helium abundance in equilibrium. Then traces of $\alpha$ nuclei remain which were not transformed into iron-group elements. 

These iron-group elements can act as seed nuclei for the weak $s$-process in future generations of stars.

\subsection{The $s$-process and $^{58}$Ni}
The $s$-process distribution in the solar system can be divided into three components: a "weak" (60$<$A$<$90 \cite{PGH10}), a "main" (90$<$A$<$208 \cite{BTG14}), and a "strong" component (mostly producing half of the solar $^{208}$Pb \cite{GAB98}), corresponding to different astrophysical scenarios, temperatures, and neutron densities \cite{KGB11}. The main and the strong $s$-process occur mainly in low- and intermediate-mass (1$-$3~M$_\odot$) "thermally pulsing asymptotic giant branch" (TP-AGB) stars at different metallicities \cite{TGA04}. 

The weak $s$-process component occurs in massive stars ($M>$8~$M_\odot$) during the core He- and shell C-burning phases. Near He exhaustion the temperature rises to about 300~MK ($kT=26$~keV) and activates the $^{22}$Ne($\alpha,n$)$^{25}$Mg reaction as main neutron source \cite{RBG91a,RBG91b,RGB93}. While the peak neutron densities reach only moderate values of about 10$^{6}$~cm$^{-3}$, this weak $s$-process phase can last several million years. 

A fraction of the $^{22}$Ne survives and is re-ignited in the following convective shell C-burning phase when new $\alpha$-particles are produced via the $^{12}$C($^{12}$C,$\alpha$)$^{20}$Ne reaction. At about 1~GK ($kT=90$~keV) the $^{22}$Ne($\alpha,n$)$^{25}$Mg reaction works more efficiently and reaches peak neutron densities of 10$^{10}$~cm$^{-3}$ for a duration of only a few years.

Starting with an iron-seed distribution produced during silicon burning of earlier generations of stars, the neutron capture on the abundant $^{58}$Ni is one of the first reactions in the weak $s$-process. An accurate knowledge of the neutron capture cross section at $s$-process temperatures is thus an important prerequisite for understanding the reaction network of the $s$-process.

\subsection{Previous measurements of the $^{58}$Ni$(n,\gamma)$$^{59}$Ni cross section at stellar energies}
The $^{58}$Ni$(n,\gamma)$$^{59}$Ni cross section has been measured at stellar energies in several time-of-flight (TOF) experiments \cite{BSE74,BeS75,WKR84,PPH93,GDL10,PZ14}. The overview in the "Karlsruhe Astrophysical Database of Nucleosynthesis in Stars" (KADoNiS) \cite{kad-58ni} (see also Table~\ref{tab:comp}) reveals that "older" measurements which were performed before 1993 \cite{BSE74,BeS75,WKR84,PPH93} yield a systematically higher Maxwellian-averaged cross section at $kT$= 30~keV (MACS30) compared to the two more recent TOF measurements from ORELA \cite{GDL10} and n$\_$TOF \cite{PZ14}. The latter two are in agreement and were used for the derivation of the new recommended MACS at $kT=30~$keV  of $\langle\sigma\rangle_{30keV}= 34.1(15)$~mbarn in the most recent release of the KADoNiS v1.0 database.

To investigate this systematic discrepancy we studied this reaction with a completely independent method. By combining the neutron activation of two Ni samples of natural isotopic composition in a quasi-stellar neutron spectrum with the subsequent atom counting of the reaction product $^{59}$Ni using accelerator mass spectrometry (AMS), the cross section of the $^{58}$Ni$(n,\gamma)^{59}$Ni reaction was extracted. The preliminary results were reported in a previous conference proceeding \cite{RDF07}. However, the first sample suffered from a relatively high isobaric contamination of $^{59}$Co. As a consequence a chemical purification step was introduced for the second sample, as described in Sec.~\ref{sec:ams}. Our preliminary result ($\langle\sigma\rangle_{30keV}=30.0(23)$~mbarn \cite{RDF07}) is in conflict with the previous TOF measurements and the new recommended MACS ($\langle\sigma\rangle_{30keV}= 34.1(15)$~mbarn). %$\langle\sigma\rangle_{30keV}=38.7(15)$~mbarn. 
This discrepancy triggered the irradiation of a third Ni sample with a lower initial $^{59}$Co content, which did not require a chemical treatment prior to the AMS measurement. 

In this paper we present the final results of all three samples, activated at KIT with a quasi-Maxwellian neutron spectrum of $kT = 25~$keV and measured for the $^{59}$Ni content at the 14~MV tandem accelerator at the Maier-Leibnitz laboratory in Garching between April 2005 and January 2011. The neutron irradiations at the (now closed) 3.7~MV Van-de-Graaff accelerator in Karlsruhe and the determination of the neutron fluence are described in Sec.~\ref{sec:n-exp}. The AMS measurements and the chemical sample preparation for the reduction of the interfering $^{59}$Co isobar are described in Sec.~\ref{sec:ams}. The derivation of the resulting MACS and a comparison with previous measurements, evaluated libraries, as well as theoretical predictions, are shown in Sec.~\ref{sec:macs}.

%%%%%%%%%%%%%%%%%%%%%%%%%%%%%%%%%%%%%%%%%%%%%%%%%%%%%
\section{Neutron activation}\label{sec:n-exp}
%%%%%%%%%%%%%%%%%%%%%%%%%%%%%%%%%%%%%%%%%%%%%%%%%%%%%
\subsection{Activation setup}
\begin{figure*}[!htb]
	\centering
		\includegraphics{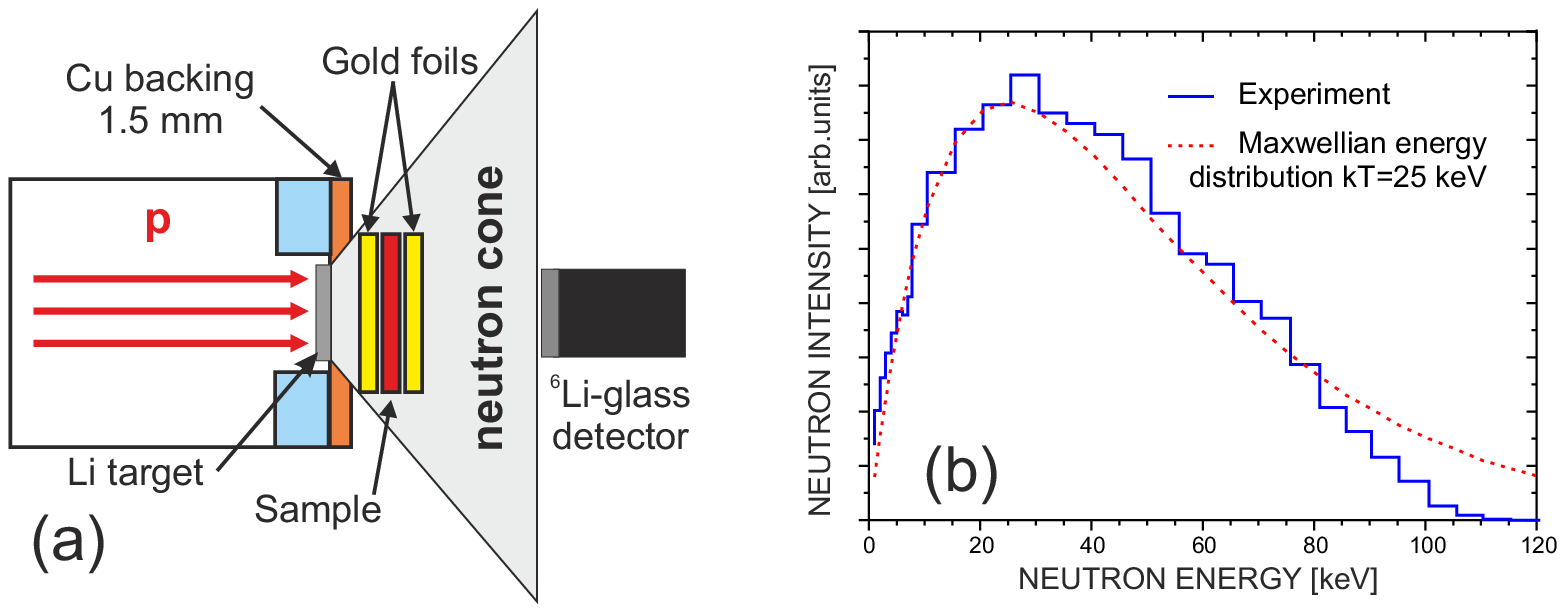}
	\caption{(Color online) a) Schematic drawing of the neutron production target and position of the samples during irradiation. At $E_p=1912$~keV the neutrons are kinematically collimated in a forward cone with 120$^\circ$ opening angle. b) Comparison of the experimental neutron spectrum with a Maxwellian energy distribution of $kT=25$~keV.}
	\label{fig:li7-spectrum}
\end{figure*}

The activation was carried out with the Karlsruhe 3.7~MV Van de Graaff accelerator. Neutrons were produced with the $^7$Li($p,n$)$^7$Be source by bombarding $20-30~\mu$m thick layers of metallic Li on a water-cooled Cu backing with protons of 1912 keV, 31~keV above the reaction threshold. The angle-integrated neutron spectrum at this energy almost perfectly imitates a Maxwell-Boltzmann distribution for $kT = 25.0\pm 0.5~$keV with a maximum neutron energy of 106~keV \cite{Rat88}, as can be seen in Fig.~\ref{fig:li7-spectrum}.

At this proton energy the neutrons are kinematically collimated in a forward cone with 120$^\circ$ opening angle. Neutron scattering by the Cu backing is negligible since the transmission is about 98\% in the energy range of interest. To ensure a homogeneous illumination of the entire surface, the proton beam with a DC current of $\sim 100~\mu$A was wobbled across the Li target. In this way a mean neutron intensity over the period of the activations of $(1.6-2.1)\times 10^9$~s$^{-1}$ was assured at the position of the samples, which were placed in close geometry to the Li target inside the neutron cone (see Fig.~\ref{fig:li7-spectrum}). A $^6$Li-glass monitor at 1~m distance from the neutron target was used to record the time-dependence of the neutron yield in intervals of 60~s as the Li target degrades during the irradiation. In this way the proper correction of the number of nuclei which decayed during the activation (factor $f_b$ in Eq.~\ref{eq:fb}) can be obtained. This correction is negligible for isotopes with very long half-lives like $^{59}$Ni but becomes important for comparably short-lived isotopes like $^{198}$Au ($t_{1/2}=2.6941(2)$~d), since the reaction $^{197}$Au$(n,\gamma)$ was used as the reference cross section for the neutron fluence determination.

Three samples prepared from metallic Ni powder with natural composition were used in our measurements. The sample material was pressed into thin pellets of 6~mm and 8~mm diameter, enclosed in 15~$\mu$m thick aluminum foil and sandwiched between two 30~$\mu$m thick gold foils of the same diameter. In this way the neutron fluence in our experimental neutron distribution (Fig.~\ref{fig:li7-spectrum}) can be determined relative to the revised neutron capture cross section of $^{197}$Au in the experimental neutron spectrum (see Sec.~\ref{sec:renorm}).

The samples "ni-1" and "ni-2" were irradiated in the same, first activation run. The sample "ni-3" was irradiated independently 3 years later. The net irradiation times were $\sim$~6~d and $\sim$~5~d, respectively, and the total neutron fluence $\Phi_{tot}$ was calculated from the $\gamma$-activity of the gold foils (see Table~\ref{tab:act}).

\subsection{Determination of the neutron flux}
The measurement of the induced $^{198}$Au activity after the irradiation was performed with a high purity germanium (HPGe) detector with a well defined measuring position at a distance of 76 (1)~mm surrounded by 10~cm lead shielding. The relative efficiency for the 411.8~keV $\gamma$-transition into the ground-state of $^{198}$Hg was determined with a set of reference sources and yielded $\varepsilon_\gamma$=0.212 (4)\%.

The total amount of produced $^{198}$Au nuclei, $N_{198}$, at the end of the irradiation can be deduced from the number of events $C(t_m)$ in the particular $\gamma$-ray line at 411.8~keV registered in the HPGe detector during the measuring time $t_m$. The factor $t_w$ corresponds to the waiting time between the end of the irradiation and the start of the activity measurement:
\begin{eqnarray}
N_{198} = \frac{C(t_m)} {\varepsilon_\gamma~ I_\gamma~
k_{\gamma}~(1-e^{-\lambda~t_m})~e^{-\lambda~t_w}} \label{eq:Z}.
\end{eqnarray}
$I_\gamma$ accounts for the relative $\gamma$ intensity per decay of the 411.8~keV transition ($I_\gamma$= 95.58(12)\%, \cite{NDS198}). For the measurement of the activated gold foils with the HPGe, the $\gamma$-ray self-absorption $k_\gamma$ has to be considered. For disk shaped samples with a thickness $d$, $k_\gamma$ can be calculated with the $\gamma$-absorption coefficients $\mu$ \cite{NIST-gamma} via Eq.~\ref{eq:kgamma}:
\begin{eqnarray}
k_\gamma= \frac{1}{d~\mu~(1-e^{d\mu})}. \label{eq:kgamma}
\end{eqnarray}
This correction factor was 0.995 for all gold foils. 

The number of produced atoms $N_{act}$ is determined by
\begin{eqnarray}
N_{act}=N_i~\langle\sigma\rangle_{exp}~\Phi_{tot}~f_b. \label{eq:act}
\end{eqnarray}
In this equation, $N_i$ is the number of sample atoms, $\langle\sigma\rangle_{exp}$ is the spectrum-averaged neutron capture cross section in the experimental neutron spectrum, and $\Phi_{tot}$ the total neutron fluence (see Table~\ref{tab:act}). The factor
\begin{eqnarray}
f_b=\frac{\int_{0}^{t_a}\phi(t)~e^{-\lambda(t_a-t)}~dt}{\int_{0}^{t_a}\phi(t)~dt}
\label{eq:fb}
\end{eqnarray}
accounts for the decay of radioactive nuclei during the irradiation time $t_{act}$ as well as for variations in the neutron flux. This factor is calculated from the neutron flux history recorded throughout the irradiation with the $^6$Li-glass detector at 1~m distance from the target. 

\begin{table}[!htb]
\caption{Summary of the sample and activation parameters. $"t_{act}"$ is the irradiation time and "$\Phi_{tot}$" the total neutron exposure. 'Diameter' refers to the diameter of the irradiated pellet. \label{tab:act}}
\renewcommand{\arraystretch}{1.2} % enlarge line spacing
\begin{ruledtabular}
\begin{tabular}{ccccc}
Sample & N($^{58}$Ni)  & Diameter  & $t_{act}$ & $\Phi_{tot}$ \\
 & ($\times 10^{20}~$atoms) & (mm) & (d) & ($\times$10$^{14}$~n) \\
\hline 
ni-1 & 1.652 & 8 & 6.0  & 8.73 \\
ni-2 & 1.649 & 8 & 6.0  & 8.50 \\
ni-3 & 5.307 & 6 & 5.0  & 9.23 \\
\end{tabular}
\end{ruledtabular}
\end{table}

The time-integrated neutron flux $\Phi_{tot} = \int \phi(t)dt$ seen by the sample (see Table~\ref{tab:act}) was determined by averaging the neutron fluences of the two gold foils enclosing the respective sample:
\begin{eqnarray}
\Phi_{tot}= \frac{N_{198}}{N_{197}~\langle\sigma\rangle_{exp}(^{197}Au)~f_b}. \label{eq:act2}
\end{eqnarray}
$\langle\sigma\rangle_{exp}$ is the experimental spectrum-averaged $^{197}$Au cross section (see discussion in the following section). 

\subsection{The new recommended $^{197}$Au$(n,\gamma)^{198}$Au cross section} \label{sec:renorm}
$^{197}$Au is commonly used as reference for neutron capture cross section measurements. However, it is only considered a standard for thermal energies ($kT$=~25.3~meV) and in the energy range between 200~keV and 2.8~MeV \cite{Car09}. Recent high-accuracy time-of-flight measurements at n\_TOF \citep{Mas10,Led11} and at GELINA \citep{Mas14} revealed a discrepancy of 5\% at $kT$=~30~keV compared to the recommended $^{197}$Au$(n,\gamma)$$^{198}$Au cross section used in the previous versions of the KADoNiS database \cite{bao00,kad03}.

This previous recommendation was based on an activation measurement performed by the Karlsruhe group, which yielded a spectrum-averaged cross section of 586(8)~mbarn for the quasi-stellar spectrum of the $^{7}$Li$(p,n)$$^{7}$Be source at $E_p$=~1912~keV (see Fig.~\ref{fig:li7-spectrum}), from which a MACS of 582(9)~mbarn at $kT=$~30~keV was derived \cite{Rat88}. The extrapolation to higher and lower energies was done with the energy dependence measured at the ORELA facility \cite{Mac75}. 

However, all recent TOF measurements \cite{Mas10,Led11,Mas14} are in perfect agreement with the latest ENDF/B-VII.1 evaluation \cite{endfb71}, and with a new activation measurement by the group in Sevilla \cite{JiB14}. Based on this consistency of new experimental data and data libraries, KADONIS v1.0 \cite{kadonis} now uses the value of 613(7)~mb for the $^{197}$Au(n,$\gamma$) cross section for $kT$=$30$~keV. For the astrophysically relevant energy region between $kT$=$5$ and 50~keV the values were derived by the weighted average of the GELINA measurement and the n$\_$TOF measurement. The uncertainty in this energy range was taken from the GELINA measurement \cite{Mas14}. For the energies between $kT$=~60$-$100~keV the average of recent evaluated libraries (JEFF-3.2 \cite{jeff32}, JENDL-4.0 \cite{jendl40}, ENDF/B-VII.1 \cite{endfb71}) was used with the uncertainty from the standard deviation given in JEFF-3.2 and ENDF/B-VII.1 (Table~\ref{tab:xs2}). 

The new reference value for the spectrum-averaged $^{197}$Au cross section of the experimental neutron distribution (Fig.~\ref{fig:li7-spectrum}) becomes $\langle\sigma\rangle_{exp}$= 632(9)~mbarn and was subsequently used for the determination of the neutron fluence in Eq.~\ref{eq:act2}.

\begin{table}[!htb]
\caption{New recommended MACS of $^{197}$Au(n,$\gamma$)$^{198}$Au from KADoNiS v1.0 \cite{kadonis} in comparison with the previously recommended values \cite{bao00,kad03}. The values given in brackets are the respective uncertainties.} \label{tab:xs2}
\begin{tabular}{cccc}
 \multicolumn{4}{c}{$^{197}$Au(n,$\gamma$)$^{198}$Au} \\
 \hline
$kT$ & $\langle\sigma\rangle_{30keV}$ (mbarn) & $\langle\sigma\rangle_{30keV}$ (mbarn) & Ratio \\
$$(keV) & KADoNiS v1.0 & KADoNiS v0.3 & $\frac{v1.0}{v0.3}$\\
\hline
5 & 2109 (20) & 2050 & 1.029\\
8 & 1487 (13) & - & - \\
10 & 1257 (10) & 1208 & 1.041\\
15 & 944 (10) & 904 & 1.044\\
20 & 782 (9) & 746 & 1.048\\
25 & 683 (8) & 648 & 1.054\\
30 & 613 (7) & 582 (9) & 1.053\\
40 & 523 (6) & 496 & 1.054\\
50 & 463 (5) & 442 & 1.048\\
60 & 425 (5) & 406 & 1.047\\
80 & 370 (4) & 356 & 1.039\\
100 & 332 (4) & 312 & 1.064\\
\hline
\hline
\end{tabular}
\end{table}

%%%%%%%%%%%%%%%%%%%%%%%%%%%%%%%%%%%%%%%%%%%%%%%%%%%%%
\section{Determination of the isotopic ratio via AMS} \label{sec:ams}
%%%%%%%%%%%%%%%%%%%%%%%%%%%%%%%%%%%%%%%%%%%%%%%%%%%%%

\subsection{Accelerator Mass Spectrometry}
Accelerator Mass Spectrometry (AMS) \cite{SYN13,KUT13} is an ultra-sensitive and ultra-selective analytical method for the detection of trace amounts (sub-ng range) of long-lived radioactive isotopes like $^{59}$Ni. AMS allows the determination of the concentration of the radioisotope relative to the ion current of a stable isotope (ideally of the same element). It is the most sensitive detection method for many radioisotopes and can reach down to isotopic ratios of 10$^{-16}$ for isotopes where complete background suppression is possible (e.g. $^{14}$C, $^{60}$Fe). AMS is able to outperform decay counting techniques in cases where the radioisotope of interest is either very long-lived or lacks suitable $\gamma$ transitions. However, one of the major challenges for AMS measurements is the suppression of (stable) isobaric interference. 

AMS usually determines the isotopic ratio of a radioactive isotope relative to one stable isotope of the same element. In our case we have determined the ratio of $^{59}$Ni versus $^{58}$Ni, $R$=$\frac{N_{59}}{N_{58}}$. The factor $f_b$ (Eq.~\ref{eq:fb}) is 1 for the long-lived $^{59}$Ni so we can rewrite Eq.~\ref{eq:act2} as
\begin{eqnarray}
\langle\sigma\rangle_{exp}(^{58}\text{Ni})= \frac{N_{59}}{N_{58}}~\cdot~\frac{1}{\Phi_{tot}}. \label{eq:act3}
\end{eqnarray}
The derived experimental spectrum-averaged cross section $\langle\sigma\rangle_{exp}$($^{58}$Ni) for the $^{58}$Ni$(n,\gamma)$$^{59}$Ni reaction is given in Table~\ref{tab:xs}.

A decade ago AMS has been successfully combined with astrophysical activation measurements, mainly for the determination of $(n,\gamma)$ cross sections for $s$-process nucleosynthesis (see, e.g. \cite{ni62,RDF07,ca40}) or for the independent measurement of actinide cross sections ($^{235,238}$U$(n,\gamma)$) to resolve discrepancies in nuclear data libraries \cite{WBB14}. But also charged-particle cross sections of astrophysical interest have been measured, e.g. the $^{25}$Mg$(p,\gamma)^{26}$Al$^{\rm g}$ and $^{40}$Ca$(\alpha,\gamma)$$^{44}$Ti cross sections \cite{al26,ti44}. An overview of cross sections measured with AMS for nuclear astrophysics so far is given in Refs.~\cite{wallner} and \cite{KT10}.
%and [G. Korschinek and T. Faestermann in "The Encyclopedia of Mass Spectrometry" Vol. 5, D. Beauchemin and D.E. Matthews eds., Elsevier 2010, p. 646]

\subsection{The GAMS setup in Garching}
\begin{figure}[!htb]
	\centering
		\includegraphics{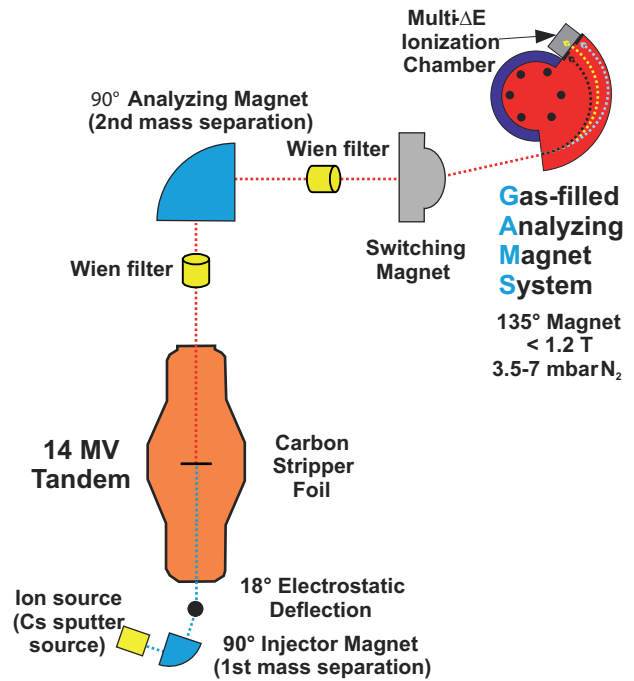}
	\caption{(Color online) Schematic illustration of the GAMS setup at the MLL in Garching (not to scale). Blue and red dotted lines represent the negatively and positively charged ion beam of interest, respectively. In the GAMS magnet, the isotope of interest follows the yellow, dotted trajectory into the detector. Black and grey dotted lines in the GAMS magnet represent isobars of higher and lower element number $Z$, respectively.}
	\label{fig:ams}
\end{figure}

The AMS setup GAMS (Gas-filled Analyzing Magnet System) at the Maier-Leibnitz-Laboratory (MLL) in Garching, Germany, is schematically shown in Fig.~\ref{fig:ams}. Its main components are a single-cathode Cs-sputter ion source, a 90$^{\circ}$ injector magnet, a 18$^{\circ}$ electrostatic deflection, a 14 MV tandem accelerator, a 90$^{\circ}$ analyzing magnet, and a Wien-filter, followed by a dedicated particle identification system. 

To separate the radioisotope  $^{59}$Ni from its stable isobar $^{59}$Co, which is orders of magnitude more abundant, the combination of a gas-filled magnet with a multi-$\Delta$E ionization chamber is employed \cite{KFK97,KFK00}. In total, the ionization chamber provides five $\Delta$E signals and the signal from the Frisch-grid (proportional to the total energy deposition in the chamber). The first two anodes are diagonally segmented, providing a horizontal position information (X-position) of the incident particles. Additionally, both, the incident X- and Y-angle, can be derived from the individual signals. 

The isotopic ratio of the radioisotope relative to a stable isotope is determined from the count-rate of the radioisotope detected in the ionization chamber, relative to the ion current of a stable reference isotope measured with a Faraday cup in front of the GAMS. Radionuclide and stable isotope were selected successively by adjusting the injector magnet, terminal voltage and Wien-filter voltage appropriately. The measurement relative to a standard sample of known isotopic ratio allows to cancel many types of systematic uncertainties such as the ion-source yield, stripping yields, and particle transmissions through the AMS system.

\subsection{Chemical treatment of the samples and preparation of the $^{59}$Ni standard}

In the first AMS beamtimes, sample ni-1 was directly pressed into sample holders without any pretreatment. This sample showed a high Co content. Therefore, the second irradiated foil (ni-2) was dissolved in 10~N hydrochloric acid (HCl) and purified by anion exchange chromatography with DOWEX AG1 resin. In a subsequent step ammonia was added until a pH value of 9 was reached, and the Ni was precipitated with dimethylglyoxim. This element-selective chelating agent was centrifuged and washed with water, and then ashed to nickel(II)oxide (NiO) which served as sample material. Due to this chemical treatment the isobar $^{59}$Co was suppressed by two orders of magnitude. The sample material for the third sample (ni-3) showed, in a first test, a much lower  $^{59}$Co background rate, thus chemical treatment was not necessary.

The $^{59}$Ni standard material '59NiO-KK92-Munich-10' was produced via irradiation of natural Ni powder with thermal neutrons. The simultaneously produced $^{65}$Ni ($t_{1/2}$= 2.52~h) served as a neutron flux monitor for the neutron fluence. The corrections for the epithermal neutron flux and sample position were on the order of 0.5\% and cancel. The activity of $^{65}$Ni was measured with a $\gamma$-ray detector utilizing the two well-known $\gamma$-ray lines at 1481.8~keV and 1115.5~keV. With the well-known thermal neutron capture cross section of $^{64}$Ni ($\sigma$= 1.52(3)~barn \cite{GK78}) and the remeasured thermal cross section of $^{58}$Ni ($\sigma$=~4.13(5)~barn \cite{ROI04}), we calculate a $^{59}\text{Ni}/\text{Ni}$ ratio for our standard of 9.1(4)$\times$10$^{-11}$. This uncertainty also includes the errors from the geometry of the sample (2\%), the statistical uncertainty of the $\gamma$-ray measurement (1.8\%) and the uncertainty of the HPGe efficiency calibration ($^{152}$Eu standard source with 2.0\% uncertainty).

\subsection{AMS procedure and data analysis for $^{59}$Ni} \label{sec:data}

\begin{figure*}[!hp]
	\centering
		\includegraphics[width=0.90\textwidth]{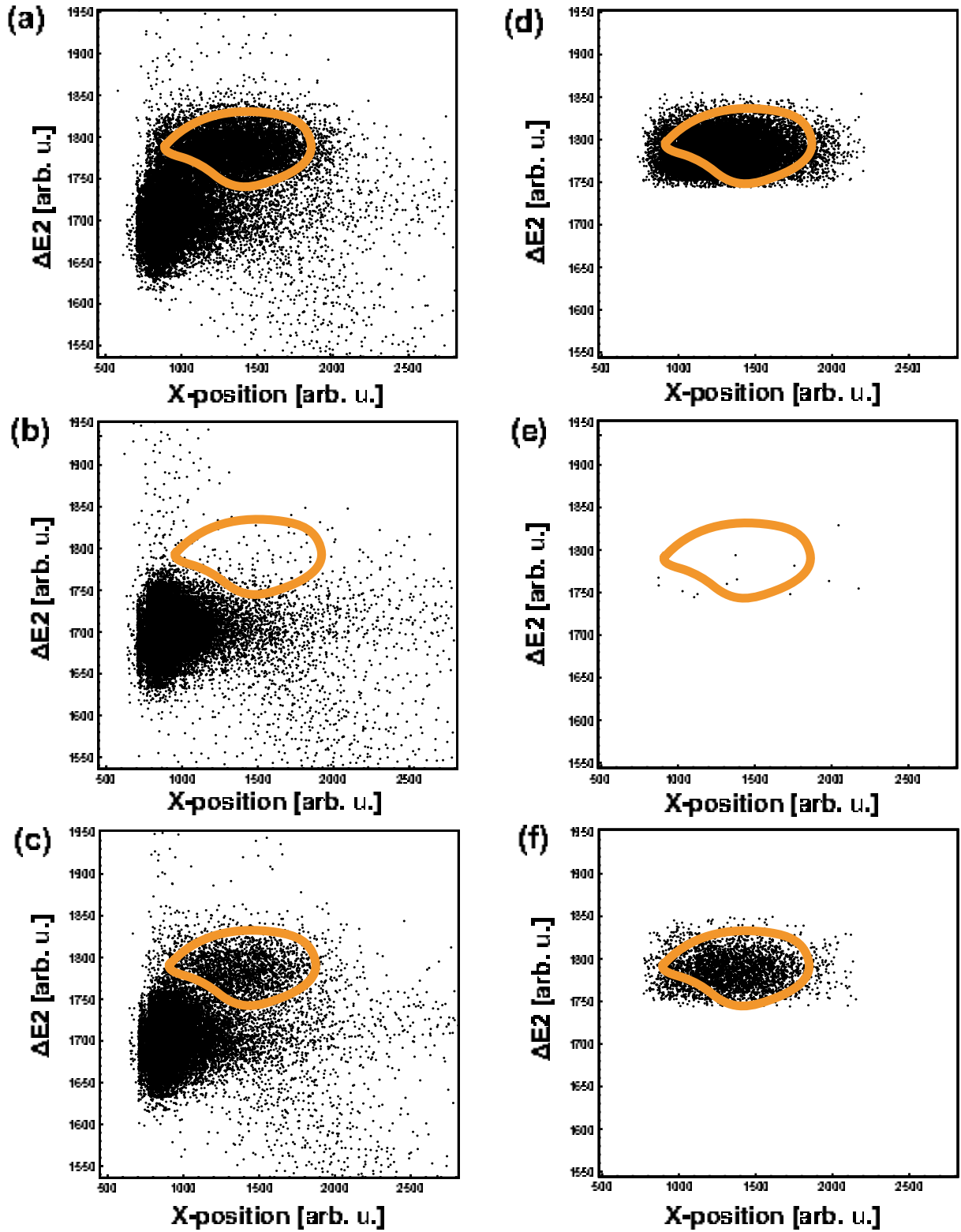}
	\caption{(Color online) Left column (a-c): Raw spectra of the energy loss in the second anode (Y-axis) versus the X-position of the incident particles (X-axis), both in arbitrary units (arb. u.), recorded with about equal statistics for (a) the standard sample, (b) a blank sample, and (c) the sample ni-2, all in linear scale in arbitrary units. Right column (d-f): the same spectra after applying a complete set of 1-dimensional software cuts around the region of interest for $^{59}$Ni on all other detector signals. Additionally, a representative example of an elliptical, 2-dimensional region of interest for $^{59}$Ni is shown in orange.The remaining background is due to the tail of $^{59}$Co (lower left of the $^{59}$Ni region).} \label{fig:spectra}
\end{figure*}

For the determination of $^{59}\text{Ni}/\text{Ni}$ ratios the tandem accelerator was operated at terminal voltages between 12.5~MV and 13.0~MV. $^{60}$Ni was used as a macroscopic beam, both for tuning of the ion optics and for normalization of the $^{59}$Ni events. The ions were extracted as $^{59}$Ni$^-$ and $^{60}$Ni$^-$ from the ion source and a typical current of 100~nA $^{60}$Ni$^-$ was injected into the accelerator. On the high-energy side of the accelerator, a charge state 12+ was selected for both isotopes with the analyzing magnet, resulting in particle energies between 162~MeV and 170~MeV for $^{59}$Ni. The GAMS magnet was filled with 5.5$-$6.7~mbar of nitrogen gas. Passing through the magnet, $^{59}$Co acquires a lower average charge state than $^{59}$Ni, due to its lower element number $Z$ (see Fig.~\ref{fig:ams}). This resulted in a spatial separation between the isobars of several centimeters horizontally and thus allowed to block most unwanted $^{59}$Co using an aperture in front of the ionization chamber. The main background in the detector was still due to tails of $^{59}$Co (see Fig.~\ref{fig:spectra}). 

In the data analysis, a region of interest for $^{59}$Ni was defined by acquiring about 1000 events of $^{59}$Ni using the standard sample. Software cuts on all signals were then applied, leading to a reduced acceptance of $^{59}$Ni of between 40\% and 60\%. The blank level, which was measured during all AMS runs using commercial Ni powder (assuming negligible $^{59}$Ni content), was always at least three orders of magnitude lower than the $^{59}\text{Ni}/\text{Ni}$ concentration of the activated and the standard samples. A final suppression of background events is achieved by applying a 2-dimensional cut on one of the energy loss signals versus the X-position of the incident ions, as shown in Fig.~\ref{fig:spectra}.

A total suppression of $^{59}$Co of about 8 orders of magnitude was achieved, combining the GAMS magnet and all software cuts. The resulting background level obtained from the commercial Ni material was $^{59}\text{Ni}/\text{Ni}=6.7\times 10^{-15}$, which was considered negligible compared to the measured ratios in our samples of $(1.7- 2.0)\times 10^{-11}$, see Table~\ref{tab:act}. 
%in of sample ni-2, which was $\frac{^{59}\text{Ni}}{\text{Ni}}$=1.7$\times 10^{-11}$.

%%%%%%%%%%%%%%%%%%%%%%%%%%%%%%%%%%%%%%%%%%%%%%%%%%%%%
\subsection{Error analysis}\label{error}
%%%%%%%%%%%%%%%%%%%%%%%%%%%%%%%%%%%%%%%%%%%%%%%%%%%%%
All uncertainties in this work are given as 1-$\sigma$ confidence intervals. The experimental uncertainties from the activation and AMS measurements are summarized in Table~\ref{tab:err1}. Note that we have separated the uncertainties into systematic contributions (which were added linearly) and statistical contributions (which were added quadratically).

\subsubsection{Uncertainties in the activation measurement}
The uncertainties from the activation measurement affect mainly the measurement of the gold standard and thus the determination of the neutron fluence $\Phi_{tot}$. The uncertainty of the $^{58}$Ni abundance can be neglected (0.03\%, see Table~\ref{tab:ni}).

%Since stellar neutron capture cross section measurements are often carried out relative to gold as standard (see discussion in Sec.~\ref{sec:renorm}), the uncertainty of 1.4\% \cite{kadonis} in the gold cross section cancels out in most astrophysical applications and was, therefore, not included in the present uncertainty analysis for the activation measurement.

The uncertainty in the mass determination of the Au foils (0.3\%) was estimated by a reading error of the used balance of $\sim$50~$\mu$g for the 20~mg foils. The neutron capture cross section of $^{197}$Au for the experimental neutron spectrum shown in Fig.~\ref{fig:spectra} has an uncertainty of 1.4\% Ref.~\cite{kadonis}. 

A random error of 1.5\% was assumed to account for the uncertainty of the sample position (0.25~mm) relative to the gold foils and the neutron target during the two activations which affects the neutron flux seen by the samples. The uncertainty of the intensity of the 411.8~keV $\gamma$-line in $^{198}$Au ($I_{\gamma}$= 95.58(12)\%) is 0.13\%, and the counting statistics yielded an uncertainty of 0.25\%. The uncertainty of the HPGe efficiency calibration was derived from the accuracy of the set of calibration sources and of the calibration procedure and yielded 2.0\%. The systematic uncertainties (including the 1.4\% from the neutron capture cross section of gold) for the activation measurement sum linearly up to 3.56\%, whereas the statistical errors yield 1.54\%. %The total uncertainty from the neutron activation was thus determined to be 3.69\%. 

\begin{table}[!htb]
\caption{Uncertainties from the activation and AMS measurement, separated as systematical and statistical contributions. Statistical contributions include regular statistical uncertainties (e.g. particle counting) and originally systematic contributions with a random effect due to repetition (e.g. sample position during irradiation)\label{tab:err1}.}
\renewcommand{\arraystretch}{1.2} % enlarge line spacing
\begin{ruledtabular}
\begin{tabular}{ccc}
  & \multicolumn{2}{c}{Activation measurement} \\
Source of uncertainty 	& Systematic (\%) & Statistical (\%) \\
\hline
$^{197}$Au cross section  & 1.4 & -- \\
Detector efficiency & 2.0 & --\\
Abundance $^{58}$Ni & 0.03 & -- \\
$\gamma$-ray intensity $^{198}$Au & 0.13 & -- \\
Sample position unc. & -- & 1.5 \\
Sample mass $^{197}$Au & -- & 0.3 \\
Counting statistics $^{198}$Au & -- & 0.25 \\
\hline
Total activation & 3.6 & 1.5 \\
\hline
& \multicolumn{2}{c}{AMS measurement} \\
Source of uncertainty & Systematic (\%) & Statistical (\%) \\
\hline
AMS standard & 4.1 & -- \\
$^{58}$Ni current\footnotemark[1] & -- & 13  \\
Beam transmission & -- & 10  \\  
Counting statistics\footnotemark[2] & -- & (5-20)  \\ 
$^{59}$Co background\footnotemark[3] & -- & 10 / - / -  \\ 
\hline
Total AMS \footnotemark[4]$^{,}$\footnotemark[5]  & 4.1 & 13.5/ 3.4/ 5.5 \\
\hline
Total act. + AMS \footnotemark[5] & 7.7 & 13.6/ 3.8/ 5.7 \\
\end{tabular}
\end{ruledtabular}
%\footnotetext[1]{Not included in the final uncertainty, see text.}
\footnotetext[1]{Per individual reading (twice per data run).}
\footnotetext[2]{Taken as $\sqrt{N}/N$ per data run with $N$ events.}
\footnotetext[3]{Only included for sample ni-1.}
\footnotetext[4]{Per data run, without counting statistics.}
\footnotetext[5]{Total statistical error for samples ni-1/ ni-2 / ni-3.}
\end{table}

\subsubsection{Uncertainties in the AMS measurement}
The AMS measurements were carried out in a 'sandwich' type order, where the actual samples of interest were measured between two measurements of the standard sample (3 data runs each), to account for instrumental drift. Using the known concentration of $^{59}$Ni/$^{60}$Ni in the standard, the transmission from the Faraday cup in front of the GAMS to the particle detector can be calculated. The typical uncertainty in transmission was 10\%, which includes current readings and counting statistics for the standard sample. 

Additionally, 13\% statistical uncertainty were included for each $^{58}$Ni ion-current reading of the sample of interest, which was determined before and after each data run (average run time 300~s). This results in a statistical uncertainty of each individual AMS data run of 13.6\%. This needs to be increased to account for the statistical uncertainty in counting statistics of $^{59}$Ni (calculated as $\sqrt{N}/N$ for typically $N=100$ events of $^{59}$Ni per data run). The results for the concentrations of $^{59}$Ni/$^{58}$Ni for each sample is then calculated as an error weighted mean from all beamtimes. A total of 20 and 10 data runs were recorded for samples ni-2 and ni-3, respectively.

An exception was made for the sample ni-1, which was only measured in three data runs in a single beamtime and suffered from a high $^{59}$Co contamination. For this sample, an additional $10\%$ contribution to the uncertainty was introduced to account for statistical fluctuations in the $^{59}$Co suppression.

Subsequently, an error weighted mean concentration of $^{59}$Ni/Ni was calculated for all data runs of each individual sample, as summarized in Tab. V.

Owing to the technique of performing AMS measurements of isotope ratios relative to a standard sample of known concentration of the radioisotope in question, all systematic uncertainties related to ion-source performance, stripping yield, and transmission cancel out. The only systematic uncertainty taken into account for the AMS result in this work is thus the uncertainty of the concentration of the standard sample (4.09\%).

%\begin{table}[!htb]
%\caption{Uncertainties from the AMS measurement, separated as systematical and statistical contributions. \label{tab:err2}}
%\renewcommand{\arraystretch}{1.2} % enlarge line spacing
%\begin{ruledtabular}
%\begin{tabular}{cccc}
%& \multicolumn{3}{c}{Final uncertainty} \\
%Sample & Data runs 	& Systematic (\%) & Statistical (\%) \\ 
%\hline
%ni-1 & 3    & 6.22 & 13.55 \\
%ni-2 & 20   & 6.22 & 3.77 \\
%ni-3 & 10   & 6.22 & 5.68 \\
%\end{tabular}
%\end{ruledtabular}
%\end{table}

\subsection{Determination of the spectrum-averaged cross section}
The experimental spectrum-averaged cross section $\langle\sigma\rangle_{exp}$ for the $^{58}$Ni($n,\gamma$)$^{59}$Ni reaction was determined according to Eq.~\ref{eq:act3} with the known values for $\Phi_{tot}$ and $N_{^{59}\text{Ni}}$/$N_{^{58}\text{Ni}}$ from Table~\ref{tab:xs}. 

The weighted spectrum-averaged cross section from the three samples was calculated to $\langle\sigma\rangle_{exp}(^{58}\rm{Ni})$= 29.9~mbarn (see Table~\ref{tab:xs}). The total systematic uncertainty (7.7\%, see Table~\ref{tab:err1}) is $\pm$2.3~mbarn, whereas the total statistical uncertainty is calculated as squareroot of the quadratic sum from the statistical uncertainty of each sample and yields $\pm$0.9~mbarn ($\pm$3.0\%). In the following sections we will quote separately the total systematical and statistical errors (e.g. for comparison with the latest measurement from n\_TOF \cite{PZ14}). %, and also the "total" uncertainty of $\pm$7.0\%.

The spectrum-averaged cross section $\langle\sigma\rangle_{exp}=~29.9~$mbarn is used in Sec.~\ref{sec:macs} for the determination of the Maxwellian-averaged cross section.

\begin{table*}[!htb]
\caption{Overview of the results from the AMS measurements and the deduced experimental cross sections $\langle\sigma\rangle_{exp}$ weighted with the statistical uncertainties from the activation and the AMS measurement. The isotope ratio is measured as $N_{^{59}\text{Ni}}$/$N_{^{60}\text{Ni}}$ and transformed into $N_{^{59}\text{Ni}}$/$N_{^{58}\text{Ni}}$ and $N_{^{59}\text{Ni}}$/$N_{\text{Ni}}$ using the corresponding isotopic abundances (see Table~\ref{tab:ni}). For discussion of the uncertainties, see Sec.~\ref{error}. \label{tab:xs}}
\renewcommand{\arraystretch}{1.2} % enlarge line spacing
\begin{ruledtabular}
\begin{tabular}{ccccccc}
Sample & AMS data runs & $N_{^{59}\text{Ni}}$/$N_{\text{Ni}}$ & $N_{^{59}\text{Ni}}$/$N_{^{58}\text{Ni}}$ & $\langle\sigma\rangle_{exp}$ & Systematic & Statistical \\
 & & & & (mbarn) & (mbarn) & (mbarn)\\
\hline 
ni-1 & 3 & 1.80$\times$10$^{-11}$ & 2.65$\times$10$^{-11}$ & 30.4 & 2.3 & 4.1  \\
ni-2 & 20 & 1.68$\times$10$^{-11}$ & 2.47$\times$10$^{-11}$ & 29.0 & 2.2 & 1.1   \\
ni-3 & 10 &  2.02$\times$10$^{-11}$ & 2.97$\times$10$^{-11}$ & 32.2 & 2.5 & 1.8  \\
\hline
\multicolumn{4}{c}{Weighted average} & 29.9 & 2.3 & 0.9  \\
\end{tabular}
\end{ruledtabular}
\end{table*}

%\begin{table}[!htb]
%\caption{Experimental cross sections $\langle\sigma\rangle_{exp}$ weighted with the statistical uncertainties from the activation and the AMS measurement.  \label{tab:xs}}
%\renewcommand{\arraystretch}{1.2} % enlarge line spacing
%\begin{ruledtabular}
%\begin{tabular}{ccccc}
%Sample & $\langle\sigma\rangle_{exp}$ & Systematic & Statistical & Total \\
% & [mbarn] & [mbarn] & [mbarn] & [mbarn] \\
%\hline 
%ni-1 & 30.35 & 1.90 & 4.11 & \\
%ni-2 & 29.04 & 1.81 & 1.09 &  \\
%ni-3 & 32.19 &  2.01 & 1.83 & \\
%\hline
%Weighted average & 29.90 & 1.87 & 0.91 & 2.71 \\
%\end{tabular}
%\end{ruledtabular}
%\end{table}

%%%%%%%%%%%%%%%%%%%%%%%%%%%%%%%%%%%%%%%%%%%%%%%%%%%%%
\section{Calculation of Maxwellian-averaged cross sections}\label{sec:macs}
%%%%%%%%%%%%%%%%%%%%%%%%%%%%%%%%%%%%%%%%%%%%%%%%%%%%%
In an astrophysical environment with temperature $T$, interacting particles are quickly thermalized by collisions in the stellar plasma, and the neutron energy distribution can be described by a Maxwell-Boltzmann spectrum:
\begin{eqnarray}
\Phi = dN/dE_n \sim \sqrt{E_n} \cdot e^{-E_n /kT} \label{eq:phi}.
\end{eqnarray}          

For the calculation of a Maxwellian-averaged cross section (MACS) from our spectrum-averaged cross section $\langle\sigma\rangle_{exp}$ we applied the following procedure: the experimental neutron spectrum of the $^7$Li($p,n$)$^7$Be reaction simulates the energy dependence of the flux $v~\cdot~\Phi~\sim~E_n~\cdot~e^{-E_n /kT}$ with $kT=25.0\pm$ 0.5~keV almost perfectly \cite{Rat88} (see Fig.~\ref{fig:li7-spectrum}). However, the cutoff at $E_n=106~$keV and small deviations from the shape of the ideal Maxwellian spectrum require a correction of the measured cross section $\langle\sigma\rangle_{exp}$ for obtaining a true Maxwellian average at $kT= 25$~keV, $\langle\sigma\rangle$$_{\rm 25keV}$. This correction factor is determined by a comparison with the energy-dependent cross sections from data libraries.

We determine a normalization factor $F_{norm}$ which gives a direct measure of the agreement between our experimentally determined cross sections $\langle\sigma\rangle_{exp}$ and the evaluated cross section $\sigma(E)$ folded with the experimental neutron spectrum of the $^7$Li$(p,n)$$^7$Be source, $\langle\sigma\rangle_{eval}$. Here we assume that the energy dependence of the library data in this energy range is correct, but the absolute scale may differ.

\subsection{Evaluated cross section libraries}\label{sec:eval}

The energy-dependent neutron cross sections from the evaluated cross section libraries ENDF/B-VII.1 \cite{endfb71}, JEFF-3.2 \cite{jeff32}, and JENDL-4.0 \cite{jendl40} were used for comparison. All three libraries include covariance data but with different uncertainties. JENDL-4.0 gives a 5.0\% uncertainty from the covariances for the MACS at $kT$= 30~keV, JEFF-3.2 gives 15.0\%, and ENDF/B-VII.1 gives 9.7\%.

The cross sections were folded with our experimental neutron distribution to derive the spectrum-averaged evaluated cross section, $\langle\sigma\rangle_{eval}$ and the respective normalization factor $F_{norm}$= $\langle\sigma\rangle_{exp}$/$\langle\sigma\rangle_{eval}$. 

\begin{table}[!htb]
\caption{Spectrum-averaged cross sections $\langle\sigma\rangle_{eval}$ from different evaluated libraries and the respective normalization factor $F_{norm}$= $\langle\sigma\rangle_{exp}$/$\langle\sigma\rangle_{eval}$.  \label{tab:NF}}
\renewcommand{\arraystretch}{1.2} % enlarge line spacing
\begin{ruledtabular}
\begin{tabular}{ccc}
Database & $\langle\sigma\rangle_{eval}$ (mbarn) & $F_{norm}$ \\
\hline 
ENDF/B-VII.1 \cite{endfb71} & 33.3 &  0.90 \\
JEFF-3.2 \cite{jeff32} & 38.6 &  0.77 \\
JENDL-4.0 \cite{jendl40} & 38.7 & 0.77 \\
\end{tabular}
\end{ruledtabular}
\end{table}

As can be seen from Table~\ref{tab:NF}, the calculated $\langle\sigma\rangle_{eval}$ cross sections do not agree for ENDF/B-VII.1 on one side, and JEFF-3.2 and JENDL-4.0 on the other side. This difference can be explained by the different resonance parameter data used to derive the resolved resonance region (RRR) and the respective statistical model which was used for the unresolved resonance region (URR) for $E_n>812$~keV.

The JENDL database uses the resonance parameters from an older technical report of Perey \textit{et al.} \cite{CP88} but not the published data of the same group in Ref.~\cite{PPH93}. The data for the URR come from the statistical model code CASTHY \cite{casthy}.

\begin{figure}[!htb]
\centering
\includegraphics[width=0.5\textwidth]{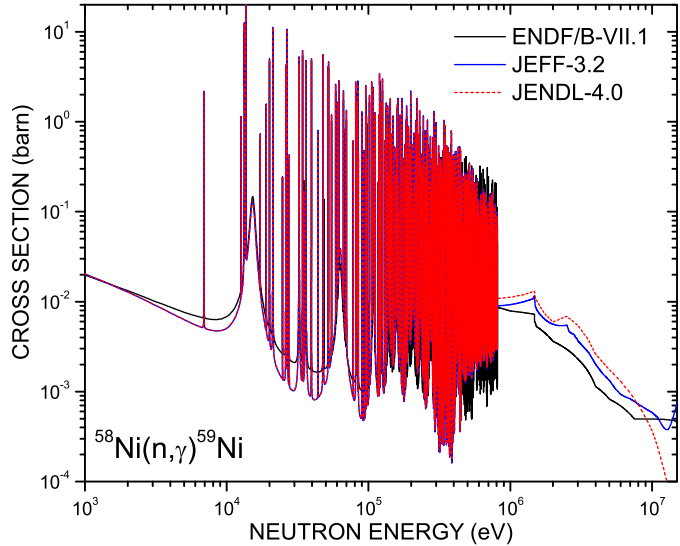}
	\caption{(Color online) Comparison of the neutron capture cross section from the evaluated libraries ENDF/B-VII.1 \cite{endfb71}, JEFF-3.2 \cite{jeff32}, and JENDL-4.0 \cite{jendl40}.}
	\label{fig:eval}
\end{figure}

JEFF-3.2 is based on older ENDF/B-VI.1 data and is thus identical with the JENDL-4.0 data in the RRR for $E_n<812$~keV. It is not obvious from the comment files in the database what was done for the URR above $E_n>812$~keV, but as can be seen in Fig.~\ref{fig:eval} the cross section is slightly lower than in the JENDL-4.0 database. However, this influence has only a minor effect on the MACS at higher energies. 

The data used in the ENDF/B-VII.1 library is the most recent and the RRR was re-evaluated by the Oak Ridge group in 2009. The following comment is given in the header of the data file: 
\textit{"The previous set of resonance parameters was based on the SAMMY analysis of ORNL neutron transmission, scattering and capture measurements by C.M Perey et al. \cite{CP88}. The present results were obtained by adding to the SAMMY experimental data base the capture cross sections recently measured at ORELA by K.H. Guber et al. (priv. comm. 2009) and the GELINA very high resolution transmission measurements performed by Brusegan et al. \cite{Brus94}. A complete Resonance Parameter Covariance Matrix (RPCM) is obtained from the SAMMY analysis of the experimental data base made consistent by neutron energy scale adjustments, and normalization and background corrections."}

The respective normalization factor to our experimental value of $\langle\sigma\rangle_{exp}(^{58}\rm{Ni})= 29.90$~mbarn is $F_{norm}=0.897$ for the ENDF/B-VII.1 library, compared to $F_{norm}=0.77$ for JEFF-3.2 and JENDL-4.0 which are based on older data. Due to this and the perfect agreement with the following publication from the ORNL group \cite{GDL10} and the independently measured data from the n\_TOF group \cite{PZ14} we used only the cross section from the ENDF/B-VII.1 library for our calculation of the MACS.

\subsection{Maxwellian-averaged cross sections}\label{sec:comp}

For the calculation of the MACS between $kT$=5$-$100~keV we have multiplied the evaluated cross section from the ENDF/B-VII.1 database \cite{endfb71} with the normalization factor $F_{norm}$= 0.90. This procedure is justified since the given covariance data in this database yields MACS with uncertainties of $\pm(11.7-9.5)\%$.

The following Table~\ref{tab:macs} shows the calculated MACS from this work in comparison to the MACS from ENDF/B-VII.1 and experimental data from recent TOF measurements \cite{PZ14,GDL10}. Our result of $\langle\sigma\rangle_{30keV}$= 30.4 (23)$^{syst}$(9)$^{stat}$~mbarn is slightly lower ($\sim 10\%$) than the two TOF measurements from \cite{PZ14,GDL10} but still agrees within the uncertainties. 

\begin{table*}[!htb]
\caption{\label{tab:macs} Maxwellian averaged cross sections
$\langle\sigma\rangle_{kT}$ (in mbarn) for $^{58}$Ni calculated with the energy dependency of ENDF/B-VII.1 ("This work"). The following two columns show the values from the two TOF measurements \cite{GDL10,PZ14}, and the last column gives the new recommended MACS from the new KADoNiS v1.0 database \cite{kad-58ni} based on these two TOF measurements.}
\begin{ruledtabular}
\renewcommand{\arraystretch}{1.1} % enlarge line spacing
\begin{tabular}{cccccc}
$kT$    & \multicolumn{5}{c}{$\langle\sigma\rangle_{kT}$ [mbarn]} \\
$$[keV] & ENDF/B-VII.1 \cite{endfb71} & This work & n\_TOF \cite{PZ14} & ORELA \cite{GDL10} & KADoNiS v1.0 \\
\hline
5  & 38.7 (45) & 34.7 & 41.3 (23)$^{syst}$(6)$^{stat}$ & 39.0 (21) & 40.0 (22) \\ 
10 & 48.1 (48) & 43.2 & 50.1 (28)$^{syst}$(7)$^{stat}$ & 48.4 (22) & 49.0 (22) \\
15 & 44.7 (43) & 40.1 & 45.9 (25)$^{syst}$(7)$^{stat}$ & 44.9 (20) & 45.3 (20) \\
20 & 40.3 (39) & 36.2 & 41.0 (22)$^{syst}$(6)$^{stat}$ & 40.5 (18) & 40.7 (18) \\
25 & 36.7 (35) & 32.9 & 37.2 (20)$^{syst}$(6)$^{stat}$ & 36.9 (16) & 37.0 (16)\\
30 & 33.9 (33) & 30.4 (23)$^{syst}$(9)$^{stat}$ & 34.2 (18)$^{syst}$(6)$^{stat}$ & 34.0 (15) & 34.1 (15) \\
40 & 29.8 (29) & 26.7 & 30.3 (15)$^{syst}$(5)$^{stat}$ & 29.9 (13) & 30.1 (15)\\
50 & 26.9 (26) & 24.2 & 27.7 (14)$^{syst}$(4)$^{stat}$ & 27.1 (12) & 27.3 (12)\\
60 & 24.8 (24) & 22.2 & 25.8 (13)$^{syst}$(3)$^{stat}$ & 24.9 (11) & 25.3 (11)\\
80 & 21.7 (21) & 19.5 & 23.2 (11)$^{syst}$(3)$^{stat}$ & 21.8 (9) & 22.3 (10)\\
100 & 19.5 (18) & 17.5 & 21.3 (10)$^{syst}$(2)$^{stat}$ & 19.6 (8) & 20.3 (9)\\
\end{tabular}
\end{ruledtabular}
\end{table*}

In Table~\ref{tab:comp} we compare our MACS at $kT$= 30~keV with previously measured values, recommended data from neutron-capture cross section compilations, evaluated libraries, and pure Hauser-Feshbach models. It should be noted that all previous measurements were performed with the TOF technique, and our measurement with the activation technique involves completely different systematic uncertainties and thus is an independent confirmation of the measured cross sections. The activation technique includes already the direct capture (DC) component which otherwise has to be inferred from theoretical predictions. The DC part calculated in Ref.~\cite{GDL10} at $kT$= 30~keV is $\langle\sigma\rangle_{kT=30keV, DC}$=1.36(34)~mbarn, whereas in the n\_TOF publication of Ref.~\cite{PZ14} no additional $s$-wave DC component was considered due to the good global fit to existing evaluated data.

\begin{table}[!htb]
\caption{Comparison of Maxwellian averaged cross sections at $kT$=30~keV from previous experiments, evaluated libraries, compilations, and theoretical predictions. The values in brackets are the respective experimental uncertainties. Theoretical values are given without error bars. \label{tab:comp}}
\renewcommand{\arraystretch}{1.2} % enlarge line spacing
\begin{ruledtabular}
\begin{tabular}{lcc}
Reference & $\langle\sigma\rangle_{30~keV}$  & Ratio to \\
 &		[mbarn]			& this work \\
\hline
\multicolumn{3}{c}{Experimental data} \\
This work  			         & 30.4 (23)$^{syst}$(9)$^{stat}$ & 1.00(7) \\
n\_TOF (2014) \cite{PZ14}    & 34.2(18)$^{syst}$(6)$^{stat}$ & 1.13(6) \\
ORELA new eval. (2010) \cite{GDL10}    & 34.0 (15) & 1.12(5) \\
ORELA (1993) \cite{PPH93}    & 40.2 (60) & 1.32(20) \\
Karlsruhe (1984) \cite{WKR84}   & 38.0 (25) & 1.25(8) \\%is it 39.0???
Karlsruhe (1974/75) \cite{BSE74,BeS75}   & 39.0 (20) & 1.28(7) \\
\hline
\multicolumn{3}{c}{Compilations} \\
KADoNiS v1.0 (2016)     & 34.1 (15) & 1.12(11) \\
KADoNiS v0.3 (2009)     & 38.7 (15) & 1.27(5) \\
KADoNiS v0.2 (2006)     & 39 (2) & 1.28(7) \\
Bao \textit{et al.} (2000) & 41 (2) & 1.35(7) \\
Beer and Winters (1992) & 43.2 (28) & 1.39(9) \\
%Allen and Macklin (19xx) &  & x$\pm$y \\
\hline
\multicolumn{3}{c}{Evaluated libraries incl. covariances} \\
ENDF/B-VII.1 \cite{endfb71}		& 33.9 (33) & 1.12(11) \\
JEFF-3.2 \cite{jeff32}		    & 40.0 (60) & 1.32(20) \\
JENDL-4.0 \cite{jendl40}		& 40.1 (20) & 1.32(7) \\
%TENDL-2014 \cite{tendl2014}	& 33.9	& \\
\hline
\multicolumn{3}{c}{Hauser-Feshbach models} \\

NON-SMOKER \cite{rau00} 	& 49	& 1.61 \\
MOST (2002)	\cite{most02} 	& 78.5	& 2.58 \\
MOST (2005)	\cite{most05} 	& 72.2	& 2.38 \\
\end{tabular}
\end{ruledtabular}
\end{table}

%%%%%%%%%%%%%%%%%%%%%%%%%%%%%%%%%%%%%%%%%%%%%%%%%%%%%
\section{Summary and Conclusions}\label{concl} 
%%%%%%%%%%%%%%%%%%%%%%%%%%%%%%%%%%%%%%%%%%%%%%%%%%%%%
We have measured the $^{58}$Ni$(n,\gamma)$$^{59}$Ni cross section for the first time with the activation technique combined with subsequent $^{59}$Ni measurements using Accelerator Mass Spectrometry. A Maxwellian-averaged cross section at $kT$=30~keV of $\langle\sigma\rangle_{30keV}$= 30.4 (23)$^{syst}$(9)$^{stat}$~mbarn could be deduced. This result is slightly lower than the two recent TOF measurements from \cite{PZ14,GDL10} but agrees within the uncertainties and confirms a lower MACS30 compared to earlier TOF measurements.

The combination of the activation technique with AMS has been proven as a powerful method to determine cross sections of isotopes with long-lived reaction products, as independent confirmation of TOF measurements.

%%%%%%%%%%%%%%%%%%%%%%%%%%%%%%%%%%%%%%%%%%%%%%%%%%%%%
\begin{acknowledgments}
%%%%%%%%%%%%%%%%%%%%%%%%%%%%%%%%%%%%%%%%%%%%%%%%%%%%%
This work was supported by the DFG Cluster of Excellence 'Origin and Structure of the Universe' (www.universe-cluster.de). We thank E.-P. Knaetsch, D. Roller and W. Seith at the Karlsruhe 3.7~MV Van-de-Graaff accelerator and the operator and workshop teams at the MLL in Garching for their help and support during the measurements. I.D. was supported by EFNUDAT and the Excellence Cluster Universe.
\end{acknowledgments}

%\bibliography{58ni}% Produces the bibliography via BibTeX.

%

\end{document}